\documentclass[]{aa}
\usepackage{natbib}
\usepackage{graphicx}
\usepackage{longtable}
\usepackage{xcolor}
\usepackage[switch]{lineno}
\bibpunct{(}{)}{;}{a}{}{,}

\newcommand{\ind}[1]{_{\mathrm{#1}}}

\newcommand\Dnu{\Delta\nu}

\newcommand\Tg{\Delta\Pi_1}
\newcommand\DPii{\Delta\Pi\ind{i}}
\newcommand\DPio{\Delta\Pi\ind{o}}
\newcommand\DPix{\Delta\Pi\ind{x}}
\newcommand\DPirad{\Delta\Pi\ind{rad}}

\newcommand\numax{\nu\ind{max}}

\newcommand\thetap{\theta_{\mathrm{p}}}

\newcommand\thetago{\theta_{\mathrm{g,o}}}
\newcommand\thetagi{\theta_{\mathrm{g,i}}}
\newcommand\thetagx{\theta_{\mathrm{g,x}}}

\newcommand\qp{q}
\newcommand\qg{q_{\mathrm{g}}}

\newcommand\np{{n\ind{p}}}

\newcommand\ngo{{n\ind{g,o}}}
\newcommand\ngi{{n\ind{g,i}}}

\newcommand\epsgo{\varepsilon\ind{g,o}}
\newcommand\epsgi{\varepsilon\ind{g,i}}
\newcommand\epsgx{\varepsilon\ind{g,x}}

\newcommand\Kepler{\textsl{Kepler}}

\def\nombre{{$41$}}

\begin{document}

\title{Extra modes in helium-core-burning stars probing an infra core cavity}
\titlerunning{Supernumerary peaks}
\author{%
B. Mosser\inst{1},
M. Takata\inst{2},
C. Pin\c con\inst{1,3},
M.S. Cunha\inst{4},
M. Vrard\inst{5,4},
K. Belkacem\inst{1},
S. Deheuvels\inst{6},
M. Matteuzzi \inst{7,8}
}

\institute{
LIRA, Observatoire de Paris, Universit\'e PSL, CNRS, Sorbonne Universit\'e, Universit\'e Paris Cit\'e, CY Cergy Paris Universit\'e, 92190 Meudon, France; \texttt{benoit.mosser@observatoiredeparis.psl.eu}
\and
Department of Astronomy, School of Science, The University of Tokyo, 7–3–1 Hongo, Bunkyo-ku, Tokyo 113–0033, Japan
\and
Universit\'e Paris-Saclay, CNRS, Institut d'Astrophysique Spatiale, 91405, Orsay, France
\and
Instituto de Astrof\'\i sica e Ci\^encias do Espaço, Universidade do Porto, CAUP, Rua das Estrelas, 4150-762 Porto, Portugal
\and
Universit\'e C\^ote d'Azur, Observatoire de la C\^ote d'Azur, CNRS, Laboratoire Lagrange, Bd de l'Observatoire, CS 34229,
06304 Nice Cedex 4, France
\and
IRAP, Universit\'e de Toulouse, CNRS, CNES, UPS, 31400 Toulouse, France
\and
Department of Physics \& Astronomy `Augusto Righi', University of Bologna, via Gobetti 93/2, 40129 Bologna, Italy
\and
INAF-Astrophysics and Space Science Observatory of Bologna, via Gobetti 93/3, 40129 Bologna, Italy
}

\abstract{Dipole mixed modes observed in the oscillation pattern of red giant stars probe the radiative regions in the stellar core.}
{Oscillation spectra of helium-core-burning stars sometimes show extra peaks that remain unexplained by the dipole mixed-mode pattern expected from the coupling of a radiative cavity in the stellar core and a pressure cavity in the stellar envelope.}
{We use the asymptotic expansion developed for a multi-cavity star in order to characterize these extra peaks.}
{The analytical resonance condition of the multi-cavity gravito-acoustic modes, with two inner gravity cavities and an outer pressure cavity, helps us explain that the apparent extra peaks are dipole mixed modes that follow the three-cavity oscillation pattern. The derivation of the two asymptotic period spacings associated with the two distinct regions in the radiative core provides an estimate of the full radiative cavity.
}
{Our results provide new constraints for analyzing the overshoot or mixing in the core of helium-core-burning stars. An important structure discontinuity inside the radiative core may explain the larger than expected observed period spacings.}


\maketitle

\section{Introduction}

Asteroseismology has proven to be extremely powerful for probing stellar interiors and constraining stellar parameters \citep[e.g.,][]{2014A&A...569A..21L}. This potential is even more developed for red giant stars, due to the ability of mixed oscillations to probe the radiative stellar core through their gravity component. Indeed, unique information can be derived from these mixed modes: transport of angular momentum \citep[e.g.,][]{2013A&A...549A..74M,2018A&A...616A..24G}, mass transfer \citep[e.g.,][]{2022A&A...659A.106D}, inner magnetic field \citep[e.g.,][]{2022Natur.610...43L}, or mixing of chemicals \citep[e.g.,][]{2015MNRAS.452..123C,2017MNRAS.469.4718B,2022NatCo..13.7553V}.

The analysis of the mixed-mode oscillation pattern largely benefits from the asymptotic formalism \citep{1979PASJ...31...87S,1989nos..book.....U,2016PASJ...68...91T}, from which seismic parameters such as the asymptotic period spacing, $\Tg$, can be derived. On the red giant branch (RGB), standard stellar modeling makes it possible to fit these spacings so precisely that outliers can be explained by a specific history such as a star merger process \citep{2022A&A...659A.106D}.
The analysis of helium-core-burning (HeCB) stars is more complicated, since observed values of $\Tg$ are significantly larger than predicted. To explain this difference, convective-core overshooting was proposed \citep{2013ApJ...766..118M,2015MNRAS.452..123C}. \cite{2017MNRAS.469.4718B} tested convective mixing as a possible reason for a smaller radiative core, and hence a larger period spacing.

Recent analyses have shown that some stars have a discrepant mixed-mode pattern with respect to the asymptotic formalism: extra peaks are sometimes present in the oscillation spectrum of HeCB stars, which cannot be explained by mixed modes resulting from the coupling of a gravity cavity and a pressure cavity \citep{2024A&A...681L..20M}. Such extra peaks were first identified in models affected by abrupt structural variations, called structural glitches \citep{2015ApJ...805..127C}. Similar extra peaks were also reported in gravity modes of subdwarf B stars \citep[e.g.,][]{2017MNRAS.472..700U}, on the hot side of the horizontal branch. So, \Kepler\ observations allow us to study these peculiar signatures on the cool side of the horizontal branch.

\begin{figure*}[t]
\includegraphics[width=18cm]{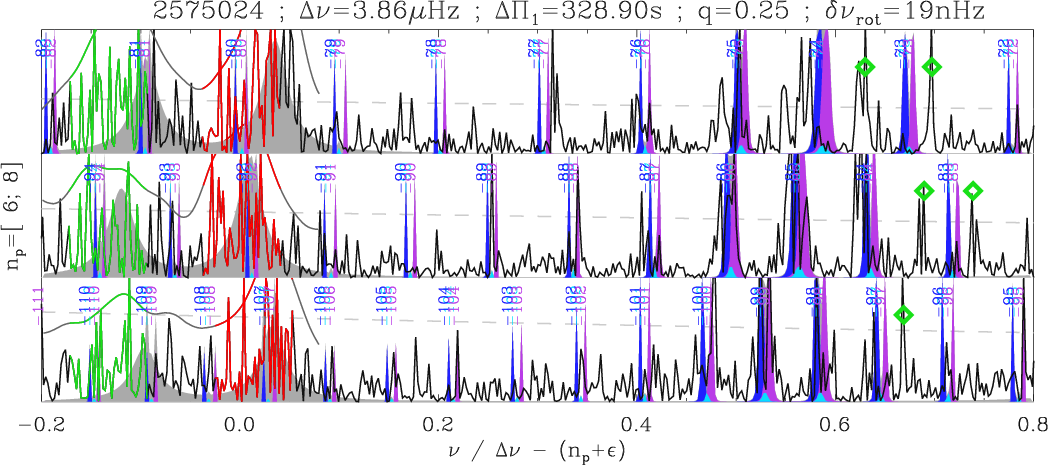}
\caption{\'Echelle spectrum of the star KIC 2575024. Radial modes, in red, and quadrupole modes, in green, appear at $x\simeq 0$ and $x\simeq -0.16$, respectively, where $x = \nu/\Dnu - (\np+\varepsilon)$.  Most of the dipole mixed modes can be fit with the two-cavity asymptotic expansion ($m=\pm1$ doublets, identified as blue and purple peaks). Extra peaks indicated by green diamonds are observed at different frequency ranges. The dashed line represents a threshold of eight times the background.
}
\label{fig-2575024}
\end{figure*}

In Section \ref{data}, we define the observed properties of the extra peaks and investigate if previous findings, such as buoyancy glitches, can account for them. We show, in Section \ref{analysis}, how the three-cavity seismic analysis depicted by  \cite{2022A&A...661A.139P} can be used to analyze these specific features. Resulting fits are analyzed in Section \ref{results} and discussed in Section \ref{discussion}.

\section{Extra peaks in the oscillation spectra\label{data}}

\Kepler\ seismic data used in this work have already benefited from multiple analyses \citep[see e.g.,][and references therein]{2025A&A...697A.165V}. We used oscillation spectra derived from 4-year time series treated according to the method of \cite{2014MNRAS.445.2698H}. Seismic analysis was performed following \cite{2018A&A...618A.109M} and \cite{2019A&A...626A.125P}, in order to provide an accurate fit of the dipole mixed-mode pattern.

From the analysis of more than 2000 HeCB stars in \cite{2024A&A...681L..20M}, we could clearly identify  supernumerary peaks in more than 10\,\% of the spectra, which induce brutal, local disruptions in the dipole mixed-mode pattern, with changes in the period spacings sometimes larger than $0.8\;\Tg$ (Fig.~\ref{fig-2575024}).
None of the usual artifacts can explain their presence. The heights and distribution of these additional peaks cannot correspond to long-lived peaks due to the presence of a binary signal superimposed on the oscillation spectrum, or to noisy peaks, or to magnetic signatures \citep{2022Natur.610...43L}.
These peaks cannot be confused with quadrupole or octupole mixed modes. They differ from the buoyancy glitch signatures depicted by \cite{2022NatCo..13.7553V}, which show significantly smaller variations and longer periods (see Appendix \ref{comparaison}).

We note that some of the extra peaks are split by rotation, similarly to classical dipole modes. This indicates their seismic origin. With this information in mind, we can consider that these extra peaks are extra dipole mixed modes.

\section{Mixed-mode analysis\label{analysis}}

Different methods can be used for analyzing the stellar oscillation spectra with extra peaks \citep{2015ApJ...805..127C,2018A&A...620A..43D,2022A&A...661A.139P}: they all account for glitches or an extra cavity, and so induce a signature that can be interpreted in terms of supernumerary peaks.
We chose to apply the three-cavity formalism of \cite{2022A&A...661A.139P}, which considers multiple cavities separated by intermediate barriers corresponding either to glitches or to evanescent regions.

\subsection{Three-cavity model}

In the three-cavity model, we considered two central g-cavities. Then, equation (112) of  \cite{2022A&A...661A.139P} reads as
\begin{equation}\label{eqt-3cavite}
  \tan \thetap (1-\qg \tan\thetago \tan\thetagi) + \qp \tan\thetago + \qp \qg \tan\thetagi = 0,
\end{equation}
where $\thetagi$ and $\thetago$ are the oscillating phases in the inner and outer g-cavity, and $\qg$ is the coupling factor between these two g-cavities. The phase, $\thetap$, and the coupling factor, $\qp$, keep the same meaning as in the two-cavity case. The phases $\thetago$  and $\thetagi$ were defined as
\begin{equation}\label{def-thetag}
  \thetagx = -\pi\left({ 1 \over \ \nu \DPix} - \epsgx \right) ,
\end{equation}
where $x$ stands for $i$ or $o$.
Six parameters were used for characterizing the pattern: the period spacings, $\DPio$ and $\DPii$, of the two g-cavities, the gravity offsets, $\epsgo$ and $\epsgi$, and the two coupling factors, $\qp$ and $\qg$.
We assumed that all these parameters are constant.

\subsection{Fitting method}

The parameters describing the outer g-cavity were first estimated from the previous analysis derived in the two-cavity case \citep{2018A&A...618A.109M}, which provides valuable proxies. Then, the three parameters of the inner g-cavity were derived from the extra peaks, identified as avoided crossings.

A first estimate of the period spacing, $\DPii$, of the inner g-cavity was derived from the period difference between two extra peaks,
\begin{equation}\label{eqt-fitDPii}
  {1\over\nu_{\ngi}} - {1\over\nu_{{\ngi}'}} \simeq (\ngi - \ngi')\ \DPii,
\end{equation}
where the indices $\ngi$ and $\ngi '$ correspond to the inner gravity radial orders of the extra peaks, defined in addition to the pressure radial order, $\np$, and the outer gravity order, $\ngo$. The challenge in this procedure is to assess the integer value $(\ngi - \ngi')$.

The frequencies of the extra peaks were also used to provide an estimate of $\epsgi$. The value of $\qg$ was derived from the smallest values of the period spacings reached in the vicinity of the extra peaks; the smaller the spacing, the smaller the coupling, as a consequence of Eq.~(\ref{eqt-3cavite}). Finally, the parameters of both g-cavities were fine-tuned iteratively.

\section{Results\label{results}}

The success of the analysis based on the three-cavity formalism relies on stringent conditions. First, at least two clear extra peaks must be identified to initiate the fitting process. Then, degenerate cases due to the location of invisible avoided crossings in frequency ranges dominated by radial or quadrupolar modes must be excluded. Difficult cases with a too low signal-to-noise ratio or too complicated rotational multiplets must be discarded. Long-period modulations reported by \cite{2022NatCo..13.7553V} are often present in addition to the supernumerary modes, which impairs the fit. As a result, not all spectra with extra peaks can be analyzed. We performed the analysis of a subset of \nombre\ stars (Fig.~\ref{fig-dP3cav}). This number is much smaller than the initial dataset, in a situation comparable to that \cite{2022NatCo..13.7553V}, who could fit long-period glitches in a number of stars much smaller than cases where a modulation is seen. Most of these fits were performed for stars where rotational triplets are reduced to a single component only (with a negligible core rotation, or seen pole on).

The coupling with pressure cavity breaks the mirror principle \citep{2003MNRAS.344..657M}: Eq.~(\ref{eqt-3cavite}) is not symmetric under the exchange between the inner and outer contributions, so we can unambiguously identify $\DPii$ as the period spacing of the inner cavity. Unsurprisingly, the values we find for $\DPio$ are close to the values of asymptotic period spacings in the two-cavity case. In fact, in the fitting process, most often we did not have to adapt the values of $\DPio$ derived from the two-cavity case: the three-cavity case can be considered as an extension of the two-cavity case. For a few stars, especially in the secondary clump, we had to slightly increase $\DPio$ compared to the original $\Tg$ value.

The values of $\DPii$, typically in the range [900,\ 2000\,s], are significantly higher than $\DPio$, so correspond to a smaller cavity. They verify in all stars the condition $\numax\ \DPii \ll 1$, which justifies the use of the asymptotic formalism.
They are also significantly smaller than the periods of buoyancy glitch found by \cite{2022NatCo..13.7553V}, typically above 4000\,s, and in line with the reduced periods measured in the cores of subdwarf B stars by \cite{2025A&A...699A.111C}.
Typical uncertainties on $\DPii$ are of about 10 - 20 s in the best case, representing a typical relative error of 1-2\,\%.  This translates into large uncertainties on $\epsgi$, since  $\delta\epsgi\simeq \delta\DPii / (\numax\DPii^2)$ (from Eq.~\ref{def-thetag}), so that no useful information can be derived from this offset.

Coupling factors, $\qg$, have small values, in the range of 2--12\,\%, smaller than the coupling $q$ between the p and g cavity, with uncertainties of a few percent. These low values of $\qg$ explain that the dipole mixed mode oscillation pattern is only locally perturbed, around extra modes, but with much larger changes than those reported by \cite{2022NatCo..13.7553V}, as a property of Eq.~(\ref{eqt-3cavite}). Far from these modes, the pattern can be depicted with the two-cavity formalism.

\begin{table}
 \caption{Seismic parameters in the three-cavity case}\label{tab-DPis}
  \begin{tabular}{lccccc}
     \hline
     KIC & $\Dnu$    & $\qp$  & $\DPio$ & $\qg$ & $\DPii$  \\
         & ($\mu$Hz) &        &  (s)    &       & (s)         \\
     \hline
  1717994&   4.11&  0.410&  318.5&  0.121& 1155.8\\
  2155220&   4.25&  0.215&  306.0&  0.038& 1896.2\\
  2448260&   4.40&  0.282&  339.5&  0.032& 1316.7\\
  2575024&   3.86&  0.208&  330.2&  0.134& 1012.8\\
  2583884&   5.51&  0.316&  307.5&  0.042& 1102.2\\
  2858440&   4.77&  0.265&  297.7&  0.044& 1344.6\\
     \hline
   \end{tabular}

The list of the full dataset with \nombre\ red-clump and secondary-clump stars is available at the CDS.
\end{table}

\begin{figure*}
\sidecaption
\includegraphics[width=14.5cm]{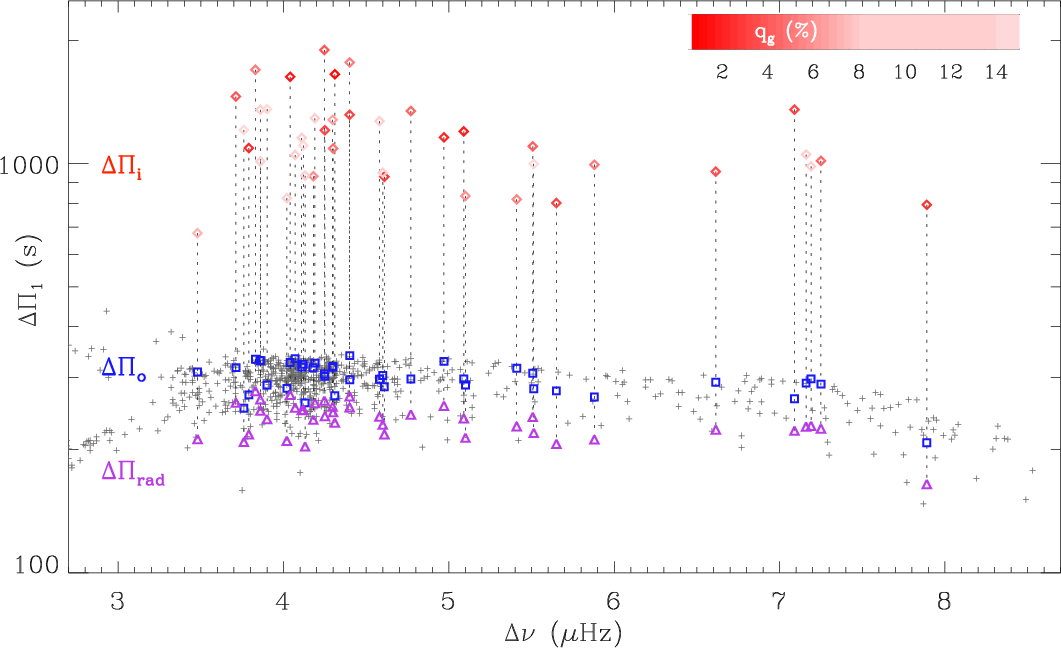}
\caption{Period spacings of the inner gravity cavity, $\DPii$ (diamonds, with a red color modulated according to $\qg$), outer gravity cavity, $\DPio$ (blue squares), and associated radiative period spacings, $\DPirad$ (purple triangles), as a function of the large separation. Asymptotic period spacings derived in the two-cavity case are also plotted (light gray crosses).
}\label{fig3cav-Dpi3cav}
\end{figure*}

\section{Radiative cavity\label{discussion}}

From the previous analysis, we can derive that the radiative core of HeCB stars is apparently split into two parts. The nature of the boundary may be a structural glitch or a region where oscillations are evanescent, since the formalism by \cite{2022A&A...661A.139P} accounts for the resonance conditions in both cases. Among phenomena that are able to split the convective core, recent studies highlight the development of semi-convection \citep{2025A&A...700A.261M, 2025A&A...704A..22P} as a possible explanation. The observed phenomenon is clearly not related to helium (sub)flashes, since their seismic signature is found both in the primary and secondary clump.

From the measurement of the two period spacings, we can derive an estimate of the extent of the total radiative cavity, expressed in terms of the global period spacing. According to the definition of the asymptotic period spacing of gravity modes applied to the inner and outer cavity \citep{1989nos..book.....U}, $\DPirad$ derives from
\begin{equation}\label{eqt-depirad}
{1\over \DPirad} \simeq {1\over \DPii} + {1\over \DPio}
.
\end{equation}
In deriving it in this way, we neglect any contribution of the intermediate region between the radiative cavities, so that we may slightly overestimate $\DPirad$. The values of $\DPirad$, smaller than $\DPio$ (Fig.~\ref{fig3cav-Dpi3cav}), seem to correspond to period spacings computed without any overshoot or mixing in the core \citep{2013ApJ...766..118M, 2015MNRAS.452..123C, 2017MNRAS.469.4718B}. This indicates that the third cavity could be the signature of the overshooting in the core, and opens a new way to analyze overshooting and mixing there. Such an analysis can also benefit from the evidence that many stars show the simultaneous signature of buoyancy glitches as depicted by \cite{2022NatCo..13.7553V} and extra peaks.

At this stage, we had not yet studied the amplitudes of the extra peaks. According to modeling \citep{2015ApJ...805..127C,2018A&A...620A..43D}, they depend on the distance from the avoided crossings located near the extra modes. For a few stars, we could identify rotational splittings. Their precise measurement is difficult, however, due to small values, but open the way to testing differential rotation in the core since they seem similar but not equal to the rotational splittings of regular mixed modes.

\section{Conclusion\label{conclusion}}

The identification of apparent extra peaks in the oscillation pattern of HeCB stars can be interpreted by the formalism proposed by \cite{2022A&A...661A.139P} accounting for resonance condition of multiple cavities. From the fits provided by the three-cavity case, we identified long-period, small coupling glitches, in a large set of HeCB stars with various masses and evolutionary stages. We infer that the radiative core is split by a structural glitch. This signature is different from the very long-period glitches reported by \cite{2022NatCo..13.7553V}, but both signatures (which sometimes appear simultaneously) may have the same origin, resulting from the  overshooting in the stellar core \citep{2013ApJ...766..118M}. On the other hand, the signature is similar to that recently studied in HeCB subdwarfs \citep{2025A&A...699A.111C}. An extensive study, beyond the scope of this work, is now necessary to analyze the full \Kepler\ set of HeCB stars.\\

Table 1 is available in electronic form at the CDS via anonymous ftp to cdsarc.u-strasbg.fr (130.79.128.5) or via http://cdsweb.u-strasbg.fr/cgi-bin/qcat?J/A+A/.

\begin{acknowledgements}
We thank the entire \emph{Kepler} team, whose efforts made these results possible.
MT was supported by the visiting fellow programme of the Paris Observatory.
MSC acknowledges support by Funda\c c\~ao para a Ci\^encia e a Tecnologia (FCT) through the  research grant UID/04434/2025 and the Investigador FCT Contract with reference 2023.09303.CEECIND/CP2839/CT0003.
MM acknowledges support from the ERC Consolidator Grant funding scheme (project ASTEROCHRONOMETRY, https://www.asterochronometry.eu, G.A. n. 772293). MV acknowledges  funding from the European Research Council (ERC) under the European Union's Horizon 2020 research and innovation programme (G.A. n. 10101965)
\end{acknowledgements}
\bibliographystyle{aa}
\bibliography{biblio_surnum}

@ARTICLE{2017MNRAS.469.4718B,
   author = {{Bossini}, D. and {Miglio}, A. and {Salaris}, M. and {Vrard}, M. and
    {Cassisi}, S. and {Mosser}, B. and {Montalb{\'a}n}, J. and {Girardi}, L. and
    {Noels}, A. and {Bressan}, A. and {Pietrinferni}, A. and {Tayar}, J.
    },
    title = "{Kepler red-clump stars in the field and in open clusters: constraints on core mixing}",
  journal = {\mnras},
archivePrefix = "arXiv",
   eprint = {1705.03077},
     year = 2017,
    month = aug,
   volume = 469,
    pages = {4718-4725},
      doi = {10.1093/mnras/stx1135},
   adsurl = {http://cdsads.u-strasbg.fr/abs/2017MNRAS.469.4718B}
}

@ARTICLE{2015MNRAS.452..123C,
       author = {{Constantino}, Thomas and {Campbell}, Simon W. and {Christensen-Dalsgaard}, J{\o}rgen and {Lattanzio}, John C. and {Stello}, Dennis},
        title = "{The treatment of mixing in core helium burning models - I. Implications for asteroseismology}",
      journal = {\mnras},
         year = 2015,
        month = sep,
       volume = {452},
       number = {1},
        pages = {123-145},
          doi = {10.1093/mnras/stv1264},
archivePrefix = {arXiv},
       eprint = {1506.01209},
 primaryClass = {astro-ph.SR},
       adsurl = {https://ui.adsabs.harvard.edu/abs/2015MNRAS.452..123C}
}

@ARTICLE{2015ApJ...805..127C,
   author = {{Cunha}, M.~S. and {Stello}, D. and {Avelino}, P.~P. and {Christensen-Dalsgaard}, J. and
    {Townsend}, R.~H.~D.},
    title = "{Structural Glitches near the Cores of Red Giants Revealed by Oscillations in g-mode Period Spacings from Stellar Models}",
  journal = {\apj},
archivePrefix = "arXiv",
   eprint = {1503.09085},
 primaryClass = "astro-ph.SR",
     year = 2015,
    month = jun,
   volume = 805,
      eid = {127},
    pages = {127},
      doi = {10.1088/0004-637X/805/2/127},
   adsurl = {http://cdsads.u-strasbg.fr/abs/2015ApJ...805..127C}
}

@ARTICLE{2025A&A...699A.111C,
       author = {{Cunha}, Margarida S. and {Amaral}, Juliana and {Avelino}, Sofia and {Falorca}, Anselmo and {Damasceno}, Yuri and {Avelino}, Pedro P.},
        title = "{Probing the cores of subdwarf B stars: How they compare to cores in helium core-burning red giants}",
      journal = {\aap},
         year = 2025,
        month = jul,
       volume = {699},
          eid = {A111},
        pages = {A111},
          doi = {10.1051/0004-6361/202553727},
archivePrefix = {arXiv},
       eprint = {2505.01381},
 primaryClass = {astro-ph.SR},
       adsurl = {https://ui.adsabs.harvard.edu/abs/2025A&A...699A.111C}
}

@ARTICLE{2018A&A...620A..43D,
       author = {{Deheuvels}, S. and {Belkacem}, K.},
        title = "{Seismic characterization of red giants going through the helium-core flash}",
      journal = {\aap},
         year = 2018,
        month = dec,
       volume = {620},
          eid = {A43},
        pages = {A43},
          doi = {10.1051/0004-6361/201833409},
archivePrefix = {arXiv},
       eprint = {1808.09458},
 primaryClass = {astro-ph.SR},
       adsurl = {https://ui.adsabs.harvard.edu/abs/2018A&A...620A..43D}
}

@ARTICLE{2022A&A...659A.106D,
       author = {{Deheuvels}, S. and {Ballot}, J. and {Gehan}, C. and {Mosser}, B.},
        title = "{Seismic signature of electron degeneracy in the core of red giants: Hints for mass transfer between close red-giant companions}",
      journal = {\aap},
         year = 2022,
        month = mar,
       volume = {659},
          eid = {A106},
        pages = {A106},
          doi = {10.1051/0004-6361/202142094},
archivePrefix = {arXiv},
       eprint = {2108.11848},
 primaryClass = {astro-ph.SR},
       adsurl = {https://ui.adsabs.harvard.edu/abs/2022A&A...659A.106D}
}

@ARTICLE{2018A&A...616A..24G,
       author = {{Gehan}, C. and {Mosser}, B. and {Michel}, E. and {Samadi}, R. and
         {Kallinger}, T.},
        title = "{Core rotation braking on the red giant branch for various mass ranges}",
      journal = {\aap},
         year = "2018",
        month = "Aug",
       volume = {616},
          eid = {A24},
        pages = {A24},
          doi = {10.1051/0004-6361/201832822},
archivePrefix = {arXiv},
       eprint = {1802.04558},
 primaryClass = {astro-ph.SR},
       adsurl = {https://ui.adsabs.harvard.edu/abs/2018A&A...616A..24G}
}

@ARTICLE{2014MNRAS.445.2698H,
       author = {{Handberg}, R. and {Lund}, M.~N.},
        title = "{Automated preparation of Kepler time series of planet hosts for asteroseismic analysis}",
      journal = {\mnras},
         year = 2014,
        month = dec,
       volume = {445},
       number = {3},
        pages = {2698-2709},
          doi = {10.1093/mnras/stu1823},
archivePrefix = {arXiv},
       eprint = {1409.1366},
 primaryClass = {astro-ph.IM},
       adsurl = {https://ui.adsabs.harvard.edu/abs/2014MNRAS.445.2698H}
}

@ARTICLE{2014A&A...569A..21L,
       author = {{Lebreton}, Y. and {Goupil}, M.~J.},
        title = "{Asteroseismology for ``{\`a} la carte'' stellar age-dating and weighing. Age and mass of the CoRoT exoplanet host HD 52265}",
      journal = {\aap},
         year = 2014,
        month = sep,
       volume = {569},
          eid = {A21},
        pages = {A21},
          doi = {10.1051/0004-6361/201423797},
archivePrefix = {arXiv},
       eprint = {1406.0652},
 primaryClass = {astro-ph.SR},
       adsurl = {https://ui.adsabs.harvard.edu/abs/2014A&A...569A..21L}
}

@ARTICLE{2022Natur.610...43L,
       author = {{Li}, Gang and {Deheuvels}, S{\'e}bastien and {Ballot}, J{\'e}r{\^o}me and {Ligni{\`e}res}, Fran{\c{c}}ois},
        title = "{Magnetic fields of 30 to 100 kG in the cores of red giant stars}",
      journal = {\nat},
         year = 2022,
        month = oct,
       volume = {610},
       number = {7930},
        pages = {43-46},
          doi = {10.1038/s41586-022-05176-0},
archivePrefix = {arXiv},
       eprint = {2208.09487},
 primaryClass = {astro-ph.SR},
       adsurl = {https://ui.adsabs.harvard.edu/abs/2022Natur.610...43L}
}

@ARTICLE{2013A&A...549A..74M,
   author = {{Marques}, J.~P. and {Goupil}, M.~J. and {Lebreton}, Y. and
    {Talon}, S. and {Palacios}, A. and {Belkacem}, K. and {Ouazzani}, R.-M. and
    {Mosser}, B. and {Moya}, A. and {Morel}, P. and {Pichon}, B. and
    {Mathis}, S. and {Zahn}, J.-P. and {Turck-Chi{\`e}ze}, S. and
    {Nghiem}, P.~A.~P.},
    title = "{Seismic diagnostics for transport of angular momentum in stars. I. Rotational splittings from the pre-main sequence to the red-giant branch}",
  journal = {\aap},
archivePrefix = "arXiv",
   eprint = {1211.1271},
 primaryClass = "astro-ph.SR",
     year = 2013,
    month = jan,
   volume = 549,
      eid = {A74},
    pages = {A74},
      doi = {10.1051/0004-6361/201220211},
   adsurl = {http://cdsads.u-strasbg.fr/abs/2013A%26A...549A..74M}
}

@ARTICLE{2025A&A...700A.261M,
       author = {{Matteuzzi}, Massimiliano and {Buldgen}, Ga{\"e}l and {Dupret}, Marc-Antoine and {Miglio}, Andrea and {Panier}, Lucy and {van Rossem}, Walter E.},
        title = "{Parametric models of core-helium-burning stars: structural glitches near the core}",
      journal = {\aap},
         year = 2025,
        month = aug,
       volume = {700},
          eid = {A261},
        pages = {A261},
          doi = {10.1051/0004-6361/202554849},
archivePrefix = {arXiv},
       eprint = {2506.18980},
 primaryClass = {astro-ph.SR},
       adsurl = {https://ui.adsabs.harvard.edu/abs/2025A&A...700A.261M}
}

@ARTICLE{2013ApJ...766..118M,
   author = {{Montalb{\'a}n}, J. and {Miglio}, A. and {Noels}, A. and {Dupret}, M.-A. and
    {Scuflaire}, R. and {Ventura}, P.},
    title = "{Testing Convective-core Overshooting Using Period Spacings of Dipole Modes in Red Giants}",
  journal = {\apj},
archivePrefix = "arXiv",
   eprint = {1302.3173},
 primaryClass = "astro-ph.SR",
     year = 2013,
    month = apr,
   volume = 766,
      eid = {118},
    pages = {118},
      doi = {10.1088/0004-637X/766/2/118},
   adsurl = {http://adsabs.harvard.edu/abs/2013ApJ...766..118M}
}

@ARTICLE{2003MNRAS.344..657M,
       author = {{Montgomery}, M.~H. and {Metcalfe}, T.~S. and {Winget}, D.~E.},
        title = "{The core/envelope symmetry in pulsating stars}",
      journal = {\mnras},
         year = 2003,
        month = sep,
       volume = {344},
       number = {2},
        pages = {657-664},
          doi = {10.1046/j.1365-8711.2003.06853.x},
archivePrefix = {arXiv},
       eprint = {astro-ph/0305601},
 primaryClass = {astro-ph},
       adsurl = {https://ui.adsabs.harvard.edu/abs/2003MNRAS.344..657M}
}

@ARTICLE{2018A&A...618A.109M,
       author = {{Mosser}, B. and {Gehan}, C. and {Belkacem}, K. and {Samadi}, R. and
         {Michel}, E. and {Goupil}, M. -J.},
        title = "{Period spacings in red giants IV. Toward a complete description of the mixed-mode pattern}",
      journal = {\aap},
         year = "2018",
        month = "Oct",
       volume = {618},
          eid = {A109},
        pages = {A109},
          doi = {10.1051/0004-6361/201832777},
       adsurl = {https://ui.adsabs.harvard.edu/abs/2018A&A...618A.109M}
}

@ARTICLE{2024A&A...681L..20M,
       author = {{Mosser}, B. and {Dr{\'e}au}, G. and {Pin{\c{c}}on}, C. and {Deheuvels}, S. and {Belkacem}, K. and {Lebreton}, Y. and {Goupil}, M. -J. and {Michel}, E.},
        title = "{Locked differential rotation in core-helium burning red giants}",
      journal = {\aap},
         year = 2024,
        month = jan,
       volume = {681},
          eid = {L20},
        pages = {L20},
          doi = {10.1051/0004-6361/202348338},
archivePrefix = {arXiv},
       eprint = {2401.07161},
 primaryClass = {astro-ph.SR},
       adsurl = {https://ui.adsabs.harvard.edu/abs/2024A&A...681L..20M}
}

@ARTICLE{2025A&A...704A..22P,
       author = {{Panier}, L. and {Buldgen}, G. and {Matteuzzi}, M. and {Scuflaire}, R. and {Dupret}, M.~A. and {Noels}, A. and {Miglio}, A.},
        title = "{Detailed theoretical analysis of core Helium-burning stars: Mixed mode patterns: I. Impact of the He-flash discontinuity and of induced semi-convection}",
      journal = {\aap},
         year = 2025,
        month = nov,
       volume = {704},
          eid = {A22},
        pages = {A22},
          doi = {10.1051/0004-6361/202555524},
archivePrefix = {arXiv},
       eprint = {2512.10474},
 primaryClass = {astro-ph.SR},
       adsurl = {https://ui.adsabs.harvard.edu/abs/2025A&A...704A..22P}
}

@ARTICLE{2019A&A...626A.125P,
       author = {{Pin{\c{c}}on}, C. and {Takata}, M. and {Mosser}, B.},
        title = "{Evolution of the gravity offset of mixed modes in RGB stars}",
      journal = {\aap},
         year = "2019",
        month = "Jun",
       volume = {626},
          eid = {A125},
        pages = {A125},
          doi = {10.1051/0004-6361/201935327},
archivePrefix = {arXiv},
       eprint = {1905.05691},
 primaryClass = {astro-ph.SR},
       adsurl = {https://ui.adsabs.harvard.edu/abs/2019A&A...626A.125P}
}

@ARTICLE{2022A&A...661A.139P,
       author = {{Pin{\c{c}}on}, C. and {Takata}, M.},
        title = "{Multi-cavity gravito-acoustic oscillation modes in stars. A general analytical resonance condition}",
      journal = {\aap},
         year = 2022,
        month = may,
       volume = {661},
          eid = {A139},
        pages = {A139},
          doi = {10.1051/0004-6361/202243157},
archivePrefix = {arXiv},
       eprint = {2203.03402},
 primaryClass = {astro-ph.SR},
       adsurl = {https://ui.adsabs.harvard.edu/abs/2022A&A...661A.139P}
}

@ARTICLE{1979PASJ...31...87S,
   author = {{Shibahashi}, H.},
    title = "{Modal Analysis of Stellar Nonradial Oscillations by an Asymptotic Method}",
  journal = {\pasj},
     year = 1979,
   volume = 31,
    pages = {87-104},
   adsurl = {http://cdsads.u-strasbg.fr/abs/1979PASJ...31...87S}
}

@ARTICLE{2016PASJ...68...91T,
   author = {{Takata}, M.},
    title = "{Physical formulation of mixed modes of stellar oscillations}",
  journal = {\pasj},
     year = 2016,
    month = dec,
   volume = 68,
      eid = {91},
    pages = {91},
      doi = {10.1093/pasj/psw093},
   adsurl = {http://cdsads.u-strasbg.fr/abs/2016PASJ...68...91T}
}

@BOOK{1989nos..book.....U,
        Address = {Tokyo},
        Adskeys = {1989nos..book.....U},
        Author = {Unno, Wasaburo and Osaki, Yoji and Ando, Hiroyasu and Saio, Hideyuki and Shibahashi, Hiromoto},
        Edition = {2nd},
        Isbn = {0860084396},
        Month = aug,
        Publisher = {University of Tokyo Press},
        Title = {Nonradial {O}scillations of {S}tars},
        Year = 1989}

@ARTICLE{2017MNRAS.472..700U,
       author = {{Uzundag}, M. and {Baran}, A.~S. and {{\O}stensen}, R.~H. and {Reed}, M.~D. and {Telting}, J.~H. and {Quick}, B.~K.},
        title = "{KIC 10001893: a pulsating sdB star with multiple trapped modes}",
      journal = {\mnras},
         year = 2017,
        month = nov,
       volume = {472},
       number = {1},
        pages = {700-707},
          doi = {10.1093/mnras/stx2011},
archivePrefix = {arXiv},
       eprint = {1812.05675},
 primaryClass = {astro-ph.SR},
       adsurl = {https://ui.adsabs.harvard.edu/abs/2017MNRAS.472..700U}
}

@ARTICLE{2022NatCo..13.7553V,
       author = {{Vrard}, Mathieu and {Cunha}, Margarida S. and {Bossini}, Diego and {Avelino}, Pedro P. and {Corsaro}, Enrico and {Mosser}, Beno{\^\i}t},
        title = "{Evidence of structural discontinuities in the inner core of red-giant stars}",
      journal = {Nature Communications},
         year = 2022,
        month = dec,
       volume = {13},
          eid = {7553},
        pages = {7553},
          doi = {10.1038/s41467-022-34986-z},
archivePrefix = {arXiv},
       eprint = {2212.11393},
 primaryClass = {astro-ph.SR},
       adsurl = {https://ui.adsabs.harvard.edu/abs/2022NatCo..13.7553V}
}

@ARTICLE{2025A&A...697A.165V,
       author = {{Vrard}, M. and {Pinsonneault}, M.~H. and {Elsworth}, Y. and {Hon}, M. and {Kallinger}, T. and {Kuszlewicz}, J. and {Mosser}, B. and {Garc{\'\i}a}, R.~A. and {Tayar}, J. and {Bennett}, R. and {Cao}, K. and {Hekker}, S. and {Loyer}, L. and {Mathur}, S. and {Stello}, D.},
        title = "{Red giant evolutionary status determination: The complete Kepler catalog}",
      journal = {\aap},
         year = 2025,
        month = may,
       volume = {697},
          eid = {A165},
        pages = {A165},
          doi = {10.1051/0004-6361/202452635},
archivePrefix = {arXiv},
       eprint = {2411.03101},
 primaryClass = {astro-ph.SR},
       adsurl = {https://ui.adsabs.harvard.edu/abs/2025A&A...697A.165V}
}

\newpage
\begin{appendix}

\onecolumn

\section{Three-cavity fit\label{3cav-fit}}

Figure \ref{fig-dP3cav} shows how the three-cavity asymptotic can fit the short period spacings due to the apparent supernumerary peaks.

\begin{figure*}[ht!]
\sidecaption
\includegraphics[width=18.5cm]{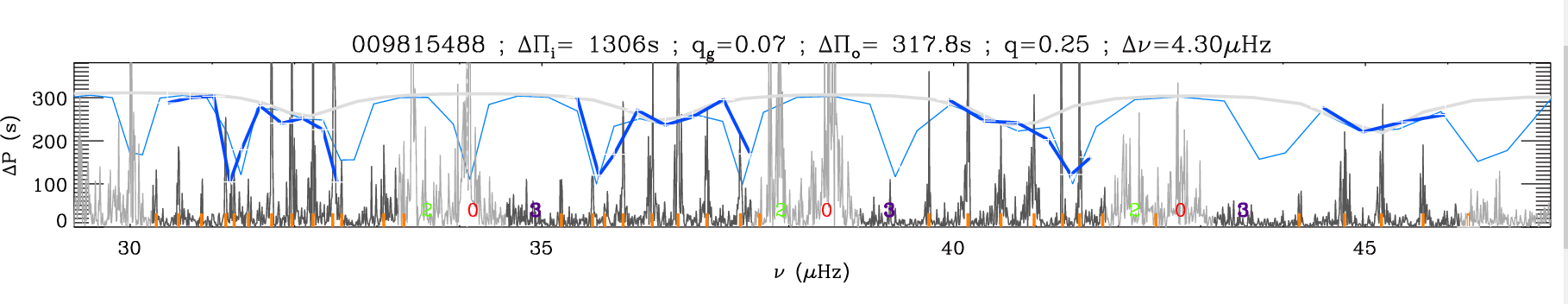}\\
\includegraphics[width=18.5cm]{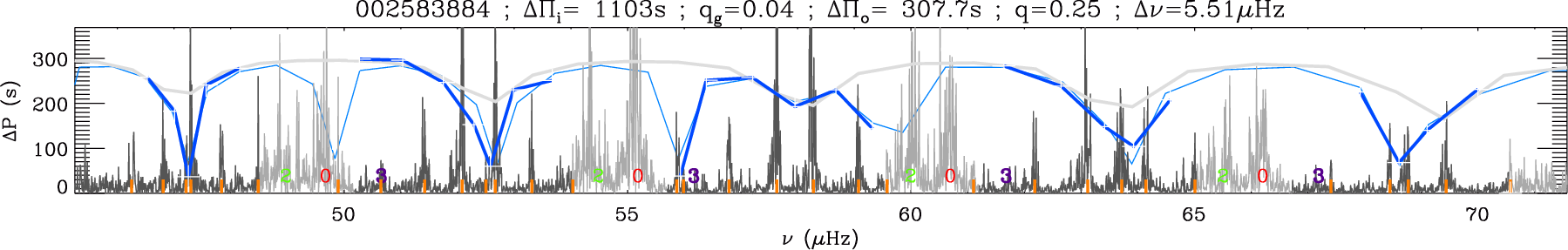}
\caption{Period spacings in the oscillation spectrum of two HeCB stars, in the red clump (KIC 9815488, top) or the secondary clump (KIC 2583884, bottom).
The gray curve indicates the expected period spacings for usual two-cavity mixed modes; the light blue curve corresponds to the three-cavity case. The dark blue curve is derived from the fitted peaks, which are identified with orange ticks. The light gray regions of the oscillation spectra indicate the ranges where radial and quadrupole modes hamper the detection of extra peaks. }\label{fig-dP3cav}
\end{figure*}

\section{Comparisons\label{comparaison}}

Different formalisms can be used to analyze the glitches \citep{2015ApJ...805..127C,2018A&A...620A..43D,2022A&A...661A.139P}.
A theoretical comparison, showing that these methods are in fact equivalent, will be performed in a forthcoming paper. Here, we propose a phenomenological illustration of their equivalence. To do so, we use the three-cavity formalism to depict the very long-period, large-coupling glitches reported by \cite{2022NatCo..13.7553V}, in two stars of their dataset that do not show apparent supernumerary modes (Figure \ref{fig-long-glitch}).

In their work, the \'echelle diagrams based on the stretched periods show a modulation with a peak-to-peak  amplitude of nearly 200\,s. Such a very long-period modulation corresponds to a much smaller variation in the period spacings shown in Fig. \ref{fig-long-glitch}. The influence of the glitch is emphasized by the difference between the two-cavity and three-cavity cases: the three-cavity period spacings show smoothed local dips.

The period spacings we extract for those stars (Table \ref{tab-DPim}) compare with those derived by  \cite{2022NatCo..13.7553V}, having in mind that the period spacings they refer to do not correspond to $\DPio$ but to $\DPirad$.  This is in line with the counting of the modes: \\
- when the inner cavity is very small ($\DPii / \DPio$ above 15, up to 50), then the apparent period spacing is close to $\DPirad$ so that the glitch signature is a modulation of this period spacing; \\
- when the inner cavity is significantly larger, as for stars we consider in our work, with  $\DPii / \DPio$ in the range [3, 6], then the glitch signature corresponds to extra narrow dips compared to the mean period spacing that is close to, but smaller than  $\DPio$. The presence of these narrow dips mimics extra modes with respect to the two-cavity case.

\begin{figure*}[t!]
\sidecaption
\includegraphics[width=18.5cm]{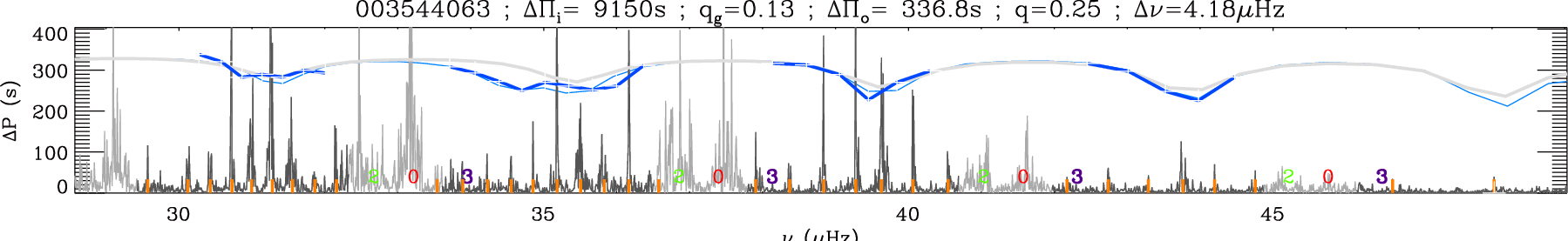}
\includegraphics[width=18.5cm]{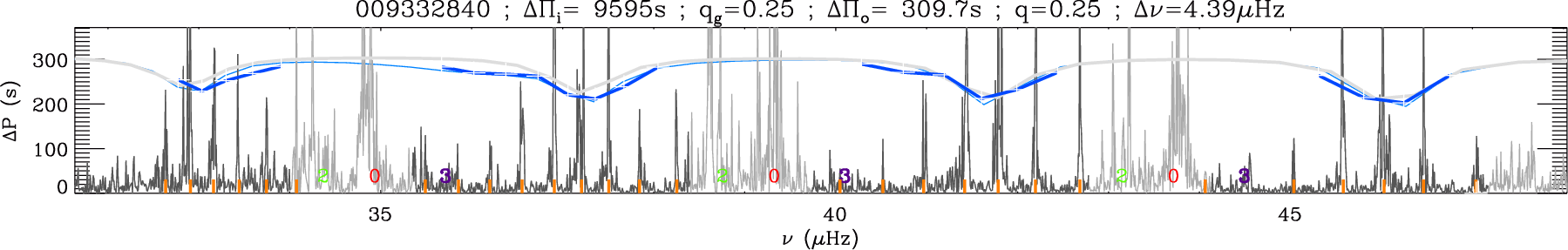}
\caption{Period spacings in the oscillation spectrum of the stars KIC 3544063 and KIC 9332840 showing evidence of structural discontinuities in the inner core \citep{2022NatCo..13.7553V}. Same styles and symbols as in Fig. \ref{fig-dP3cav}. For both stars, the glitch signature induces shorter period spacings in the frequency range around 35.5\,$\mu$Hz.}\label{fig-long-glitch}
\end{figure*}

\begin{table}[h]
 \caption{Very long-period glitches}\label{tab-DPim}
  \begin{tabular}{lccccc}
     \hline
     KIC & $\Dnu$    & $\qp$  & $\DPio$ & $\qg$ & $\DPii$  \\
         & ($\mu$Hz) &        &  (s)    &       & (s)         \\
     \hline
    3544063&  4.18 &  0.29 &   336.8 &  0.13 & 9150 \\
    9332840&  4.39 &  0.25 &   309.7 &  0.19 & 9595 \\
     \hline
   \end{tabular}

 Analysis of two stars mentioned in \cite{2022NatCo..13.7553V}.
\end{table}

\end{appendix}

\end{document}